\documentclass[aps,prx,reprint,superscriptaddress,longbibliography, floatfix]{revtex4-2}
\usepackage[T1]{fontenc}
\usepackage[english]{babel}
\usepackage{amsmath,amssymb}
\usepackage{hyperref}
\usepackage[margin=0.5in]{geometry}
\usepackage{booktabs}
\usepackage[utf8]{inputenc}
\usepackage{graphicx}
\usepackage{enumitem}
\usepackage{makecell}
\usepackage{adjustbox}	
\usepackage{bm}
\usepackage{comment}
\usepackage{booktabs}
\usepackage{tikz}
\usetikzlibrary{arrows.meta, positioning}
\usepackage{array}
\usepackage[noindentafter]
{titlesec}
\titleformat{\section}[hang]
  {\normalfont\bfseries} 
  {\thesection.}    
  {0.5em}                
  {}                     
  []                    
\titleformat{\subsection}[runin]
  {\normalfont\bfseries} 
  {\thesubsection.}    
  {0.5em}                
  {}                     
  [:]                    
\titleformat{\subsubsection}[runin]
  {\normalfont\bfseries} 
  {\thesection.\thesubsubsection}    
  {0.5em}                
  {}                     
  [:]                    

\begin{document}

\title{A proof-of-concept for automated AI-driven stellarator coil\\ optimization with in-the-loop finite-element calculations}

\author{Alan A. Kaptanoglu}
\affiliation{Courant Institute of Mathematical Sciences, New York University, New York, NY 10012, USA}
\author{Pedro F. Gil}
\affiliation{Max-Planck-Institut für Plasmaphysik, 85748 Garching, Germany}


\begin{abstract}
Finding feasible coils for stellarator fusion devices is a critical challenge of realizing this concept for future power plants. 
Years of research work can be put into the design of even a single reactor-scale stellarator design. To rapidly speed up and automate the workflow of designing stellarator coils, we have designed an end-to-end ``runner'' for performing stellarator coil optimization. The entirety of pre and post-processing steps have been automated; the user specifies only a few basic input parameters, and final coil solutions are updated on an open-source leaderboard. Two policies are available for performing non-stop automated coil optimizations through a genetic algorithm or a context-aware LLM. Lastly, we construct a novel in-the-loop optimization of Von Mises stresses in the coils, opening up important future capabilities for in-the-loop finite-element calculations.
\\
\footnotesize{
\textbf{Keywords: stellarator, coil optimization, machine learning, Large Language Model, LLM, AI}}
\end{abstract}

\maketitle


\section{Introduction}
Digital twins are increasingly seen as a critical technology towards achieving future nuclear fusion reactors~\cite{RAUSCHER2021112399,battye_digital_2025}. These tools rely on interconnected and sophisticated simulation codes for modeling different parts of the device. Moreover, such tools are likely most useful when they can be run and optimized autonomously, as there may be many hyperparameters within each submodule and substantial expertise may be required for running many of the codes. These difficulties may be especially pronounced for the stellarator, a toroidal magnetic confinement device and a promising candidate for achieving controlled nuclear fusion. Stellarators exhibit a vast parameter space that can be used for optimizing the physics and engineering performance of a reactor-scale device.


Traditionally, stellarator design is subdivided into a stage I, where the plasma boundary shape is optimized, and stage II, where the coil shapes and currents are optimized. 
This two-stage process allows for stage-1 optimization to explore suitable physics properties for fusion reactors by solving the fixed-boundary magnetohydrodynamic equilibrium equations~\cite{vmec_paper,NUHRENBERG1988113,LandremanPaul,desc_code}, while a separate stage-II optimization  focuses on finding a corresponding coilset. However, the optimization of stellarator coils is an ill-posed inverse problem; a multitude of distinct coil configurations can generate similar magnetic fields, yet most of these solutions are impractical from an engineering standpoint~\cite{stellarator_intro}. Efficiently and robustly finding accurate stellarator coils with advantageous engineering properties remains quite challenging despite substantial work in the field~\cite{Wechsung_2022, Giuliani_2024, Wu_2025,Jorge_2024,SUZUKI2021112843,Fu_2025,drevlak_onset,Pomphrey_2001,Strickler01032002,coilopt++}.
Various stellarator equilibrium concepts are now being investigated also in the industry as reactor-scale candidates~\cite{LION2025114868,Anderson_Canik_Hegna_Mowry_2025,Gates_2025, Volpe2023Renaissance, helical_fusion}, and coil complexity can have very substantial effects on the feasibility and cost margins. 

\subsection{Contributions of this work}\label{sec:contributions}
The goal of this work is to implement a series of proof-of-principle steps towards fully automated stellarator coil optimization with in-the-loop finite-element capabilities. The code can be found at \href{https://github.com/akaptano/stellcoilbench}{https://github.com/akaptano/stellcoilbench}.
%
We:
\begin{itemize}[leftmargin=*]
\itemsep0em 
    \item Automate the initialization and post-processing steps for stellarator coil optimization. 
    \item To our knowledge, we demonstrate the first in-the-loop stellarator coil optimization of Von Mises stresses.
    \item Set up a non-stop ``runner'' for performing stellarator coil optimizations.
    \item Equip that runner either with a deterministic genetic algorithm for choosing the next $N$ runs, or a context-aware Large Language Model (LLM)~\cite{chang2024survey} with access to a body of previous coil optimization solutions and manuscript summaries from the field of stellarator physics. 
    \item Demonstrate that, without supervision, the automated runs with these policies can create highly competitive coilsets on a well-studied quasi-axisymmetric plasma surface. We have found similar performance on additional plasma surfaces but have not yet performed large-scale scans. 
    \item Generate an open-source online leaderboard for each plasma surface. Any Github user can compete on the leaderboard by submitting their own \texttt{case.yaml} files for generating coil solutions or testing a new plasma design by adding an \texttt{input.plasma\_surface} file. This setup guarantees that only apples-to-apples comparisons are made using the exact same code functionality, plasma surface, plasma and coil resolution parameters, and so forth.
\end{itemize}
Most of the code for this work was generated by LLMs through the Cursor app. Extensive documentation and unit tests have been added to verify the accuracy of the code, and the final coil results of this work illustrate very high performance (and cannot be faked by the AI, as the final geometry of the coils is exported to external, well-benchmarked tools and compared with previous solutions). Any Github user can submit a properly-initialized coil optimization script or plasma surface file for competing on the leaderboard. For adding new code functionality to the code, users currently must submit an issue or pull request, and these submissions will in the future be handled by AI agents via Cursor Bugbot or similar tools. Any additional tools available through SIMSOPT~\cite{landreman_simsopt_2021} forks and branches, such as winding surface methods~\cite{landreman_improved_2017}, can be straightforwardly added. Lastly, we emphasize that this runner could be edited or repurposed to automate the runs of other coil optimization codes such as DESC~\cite{desc_code} and FOCUS~\cite{zhu_designing_2018}, or extended to run stage-1 plasma surface optimizations or single-stage optimizations.

\begin{table*}[t]
	\begin{tabular}{|l|l|l|}
		\hline
		\textbf{Component}       & \textbf{Motivation}         & \textbf{Outcome}                       \\ \hline
		\makecell{Automated\\coil initialization} & \makecell{Manual coil initialization\\ is usually required for different plasma surfaces.} & \makecell{Circular coils are initialized without coil-coil\\ or coil-surface intersections, and\\ modular coils always interlink the plasma.}      \\ \hline
		\makecell{Automated\\coil optimization} & \makecell{Manual weight tuning is expensive and\\ very different weights are needed for different plasmas.} & \makecell{Objective weights are no longer user-defined;\\ they are auto-tuned with a penalty method.\\All SIMSOPT functionality is available.}                             \\ \hline
\makecell{Genetic or\\ LLM-driven policy} & \makecell{Learning the optimal hyperparameters such as\\ thresholds, convergence tolerances, number of \\iterations, ``schedules'', etc. is very challenging.} & \makecell{Genetic algorithm or context-aware LLM\\ makes the decisions about which runs to submit\\every batch. Can keep a running log of justifications.}                             \\ \hline
    \makecell{Automated\\post-processing} & \makecell{Post-processing requires installing a host\\ of complex simulation codes and substantial training\\ of researchers to accurately use the codes.} & \makecell{Users can run new cases and plasma surfaces\\on cloud devices (through the continuous integration) where\\all post-processing is auto-installed and auto-run\\} 
    \\ \hline
		\makecell{Automated\\ documentation\\\& leaderboard} & \makecell{Historically, it has been challenging to make precise\\ comparisons between stellarator coil solutions.} & \makecell{Runs receive an overall composite score by a weighted sum\\ of coil objectives. Runs are added in real-time\\ to an open-source, online ReadTheDocs leaderboard. }
        \\ \hline
		Reproducibility &    \makecell{Historically, it has been challenging to make precise\\ comparisons between stellarator coil solutions.}
    &    \makecell{Plasma surfaces and other parameters are standardized\\and cannot be altered. Full metadata, all final solutions,\\and all metrics are saved in a database.}
        \\ \hline
	\end{tabular}
\caption{Workflow components for the automated code runner.}
\label{tab:bounds_table}
\end{table*}

\section{Components of the workflow}\label{sec:methodology}
\begin{figure*}[t]
    \centering
    \includegraphics[width=0.8\linewidth]{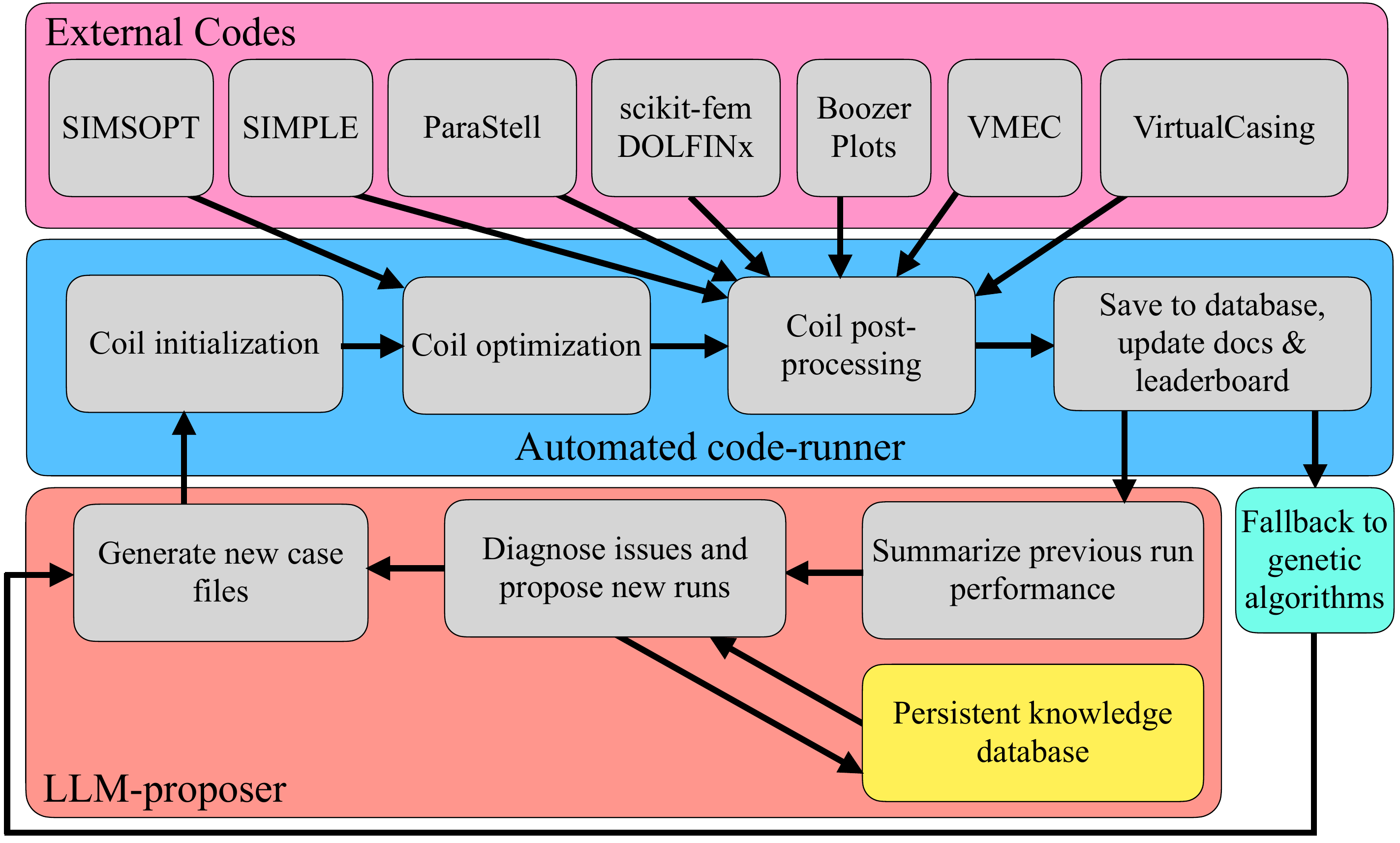}
    \caption{Automated code runner interactions with external codes and the genetic or LLM-proposer.}
    \label{fig:workflow}
\end{figure*}

Each of the components of this work are now described and further motivated in Table~\ref{tab:bounds_table}. The full workflow is summarized in Fig.~\ref{fig:workflow}.
\subsection{Automated initialization}\label{sec:initialization}
Initializing coils is an important part of stellarator coil optimization. Coil optimization is a highly nonconvex problem that can be quite sensitive to the initial condition. Much of the possible solution space is undesirable, e.g. interlinking coils are highly undesirable and coil-surface intersections are unphysical. Initial modular coils should be linked a single time with the plasma, and have significant minimum coil-coil and coil-surface distances to avoid accidental interlinking. In practice, simple circular coils are initialized, with their major and minor radii often manually tuned to fit well with the given plasma surface. The total current in the coils must also be chosen to produce the desired magnetic field on axis.

In contrast, we write an initialization loop that initializes circular coils without any manual tuning and guarantees that the modular coils avoid interlinking or intersections. Afterwards, another automated loop tunes the total current in the coils to match the desired average magnetic field. 

Moreover, the user does not need to specify any weights for the multi-objective optimization. Penalty methods (primarily with the augmented Lagrangian functionality~\cite{gil_augmented_2025}) are used by default. Thresholds for minimum coil-coil distance and other metrics are pre-defined as reasonable estimates for a reactor-scale design. For plasma surfaces that are table-top or university-scale, all of the initial weights and thresholds are appropriately scaled and nondimensionalized. We have verified that these automated loops and rescaling produce reasonable initial and final coil solutions with dozens of plasma surfaces; a systematic study is left for future work.

\subsection{Automated post-processing}\label{sec:post-processing}
Substantial post-processing is required to validate that a given coil solution sufficiently reproduces the physical properties of the plasma surface and sufficiently adheres to various engineering requirements. For this, we combine a host of well-benchmarked codes and calculations already in use in the community. For instance: 

\begin{itemize}[leftmargin=*]
\itemsep0em
\item Shape gradients~\cite{landreman2018computing} are computed on the final coils for characterizing the local coil sensitivity.
\item A direct sensitivity analysis is performed on the final coils using stochastic perturbations~\cite{Wechsung_2022,baillod2026update} of the final coilset.
    \item If there are substantial plasma currents, virtual casing~\cite{hanson_virtual-casing_2015,malhotra_efficient_2019} is used to obtain the total magnetic field from both the coils and plasma. 
    \item Poincaré plots are generated for validating that desired flux surfaces and islands are intact. 
    \item Either a free-boundary equilibrium solve can be performed, or a quadratic-flux-minimizing (QFM) surface~\cite{hudson_are_2009} must be computed, followed by a fixed-boundary solve in VMEC~\cite{vmec_paper,hirshman_preconditioned_1991} (GVEC~\cite{hindenlang_computing_2025} or DESC~\cite{desc_code,panici_desc_2023,conlin_desc_2023,dudt_desc_2023} could potentially be used too, but these codes are not yet directly supported in SIMSOPT).     Profiles of quasisymmetry and rotational transform are computed after the equilibrium solve. 
    \item Boozer coordinates~\cite{boozer_plasma_1981} using the booz\_xform package are used to visually validate quasi-symmetry. 
    \item The SIMPLE~\cite{albert_symplectic_2020} fast-particle tracing code is used here to quantify the fast-ion losses in a reactor-scale design. 
    \item The high-temperature superconductor (HTS) winding-pack model from the conceptual design for a fusion power plant, Stellaris~\cite{lion_stellaris_2025}, is used here to compute per-coil turn counts from both force and critical-current-density criteria, winding pack widths, and finite-build clearances.
    \item Meshing tools such as Gmsh~\cite{geuzaine2009gmsh} are utilized through ParaStell~\cite{moreno_parastell_2024} in order to represent finite-build coils, validate the final minimum coil-coil distances, and set-up meshes for finite-element structural and thermal analyses.
    \item Finite-element method (FEM) calculations of the stress tensor and Von Mises stresses are performed via the open-source scikit-fem~\cite{gustafsson_scikit-fem_2020} or DOLFINx package~\cite{baratta_dolfinx_2025} (for large and more sophisticated problems requiring e.g. MPI speedup). 
\end{itemize}
We emphasize that all tools that are parallelized are run as such; MPI can be used for Poincaré plotting, VMEC, and DOLFINx, and the code switches to OpenMP parallelization while running SIMPLE. Note that this is not a definitive list of functionalities, and new packages can be added.

\subsection{Policy algorithms}\label{sec:policy}
The goal of this work is to show proof-of-principle methods that can help to fully automate stellarator coil optimization. To obtain the best possible solutions, an algorithmic policy is required to learn from poor runs, improve on good runs, and explore promising parameter spaces. 

The proposer system supports two modes: (a) a deterministic genetic algorithm (mutation with log-normal threshold jitter + exploration with log-uniform sampling)~\cite{beyer_evolution_2002}, and (b) an LLM proposer with several forms of context knowledge (in the spirit of similar work in automated scientific LLMs~\cite{boiko_autonomous_2023,romera-paredes_mathematical_2024}). 
%
%
%
%
%
Policy parameters are specified that control the run batch size, context limits, the number of post-mortems to do, how much exploration vs mutation and other hyperparameters.
The automated loop runs continuously online, proposing batches, executing them, updating the leaderboard, and repeating.
%
The LLM has access to a structured knowledge base of previous optimization results, failure statistics, and domain papers, enabling it to make informed decisions about which parameter spaces to explore next.
%
Every detail of the final coils can be reproduced from the stored \texttt{coils.json} file and every optimization run can be entirely reproduced by reusing the input file \texttt{case.yaml}. Additional metadata is saved as well, e.g. the user's username, the total time taken, the hardware, and so forth.



\begin{figure*}[t]
    \centering
    \includegraphics[width=\linewidth]{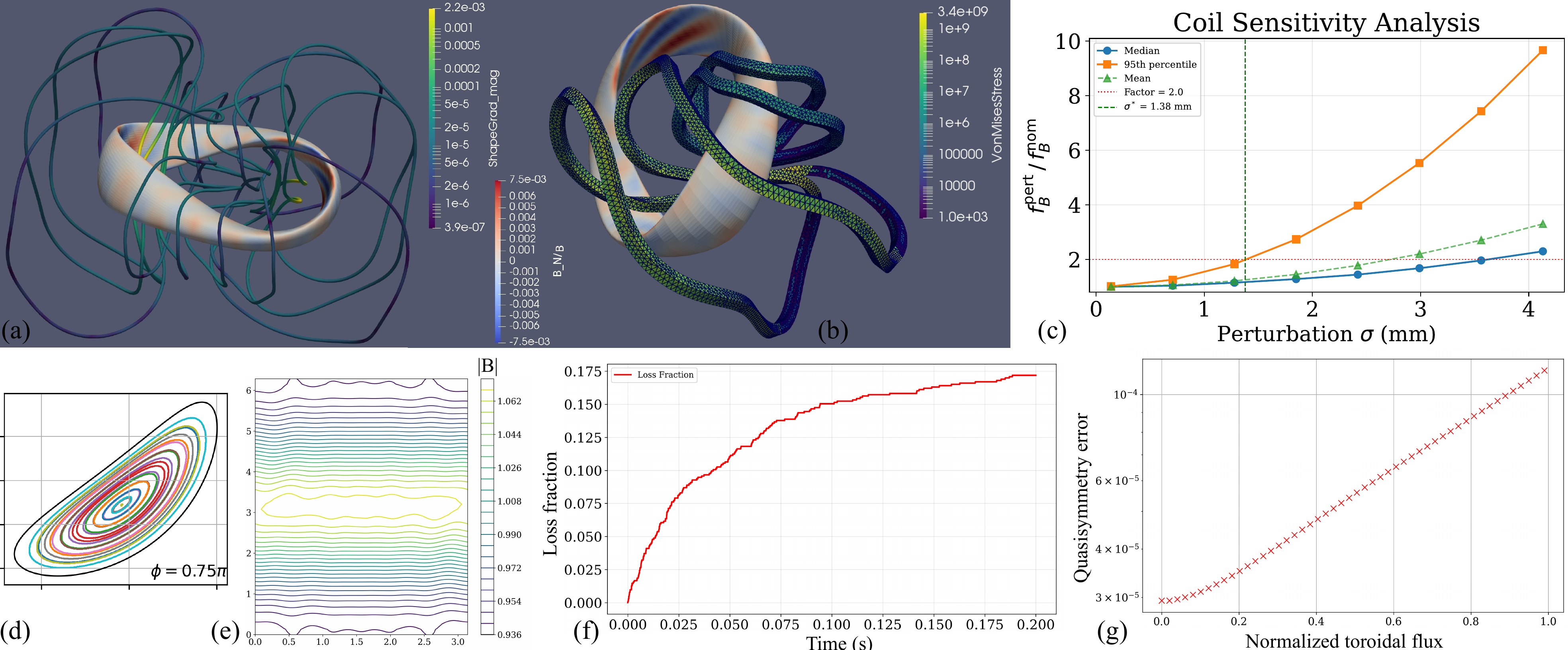}
    \caption{Result of post-processing routines. (a) Shape gradients evaluated along final coils; (b) finite-build and Von Mises stresses; (c) coil sensitivity analysis; (d) sample Poincaré plot; (e) sample Boozer plot at $s = 0$; (f) loss fraction plot; (g) two-term quasisymmetry profile.}
    \label{fig:QA_postprocessing}
\end{figure*}

\subsection{Illustration of the full pipeline}\label{sec:pipeline_illustration}
Before implementing an automated policy program, we finalize the discussion here by showing in Fig.~\ref{fig:QA_postprocessing} a manually run submission for the Landreman-Paul QA stellarator~\cite{LandremanPaul}. The run was performed with: the Augmented Lagrangian algorithm~\cite{gil_augmented_2025} (so none of the objective weights need to be specified) using the default Fourier continuation scheme in the code (Fourier orders $[4, 8, 16]$) the default objective thresholds taken from reactor-scale requirements, and the default coil optimization objectives (penalties for the total coil length, max coil curvature, mean-squared curvature, arclength variation, and Gauss linking number; precise definitions can be found many other places~\cite{zhu_designing_2018,Wiedman_Buller_Landreman_2024,hurwitz_electromagnetic_2025}). The default objectives and thresholds used in the Stellcoilbench code are tabulated in Appendix~\ref{sec:appendix_A} for completeness. The results in this section and in Sec.~\ref{sec:von_mises} use a major radius of 1 meter, toroidally-averaged $B_0 = 1$ T at the major radius, and assume square cross-section coils of width $0.1$ m. This run is by no means optimal. We use it here to demonstrate that the default functionality produces fairly good baseline coils, and to demonstrate the available post-processing tools.

Figure~\ref{fig:QA_postprocessing} illustrates most of the available options for post-processing the coil solution. Some interesting conclusions arise from the post-processing. From: (a) the shape sensitivity is exponentially larger for the parts of the coils in the inboard side of the highly curved part of the plasma surface, (b) high Von Mises stresses are found (although note that this is somewhat an artifact of having small-width finite build coils in a high-field stellarator of $1$ m major radius) (c) coil sensitivity analysis shows that on average, total $f_B$ errors degrade by a factor of 2 with just a 1.38 mm perturbation to the coils, (f-g) quasi-symmetry remains very low but there are significant enough errors to dramatically increase the loss fraction. These findings affirm previous results e.g. in Gil et al.~\cite{gil_augmented_2025} and elsewhere that some plasma surfaces can be extremely sensitive to small errors on the plasma boundary.

\section{In-the-loop optimization of Von Mises stresses}\label{sec:von_mises}
Calculation of the coil displacements and Von Mises stresses is one of the most important feasibility checks on a final coil design. Coil deformations on the order of centimeters are common from the presence of enormous electromagnetic forces~\cite{bykov2009structural}; in the best case scenario, the deformations only change the magnetic field from the coils, and this effect alone can very significantly affect the plasma operation. However, it is expected that the full effect of electromagnetic forces also impacts the overall structural integrity of a device \cite{bykov_2019},  and large Von Mises stresses can cause material failures.

\begin{figure}
    \centering
    \includegraphics[width=0.99\linewidth]{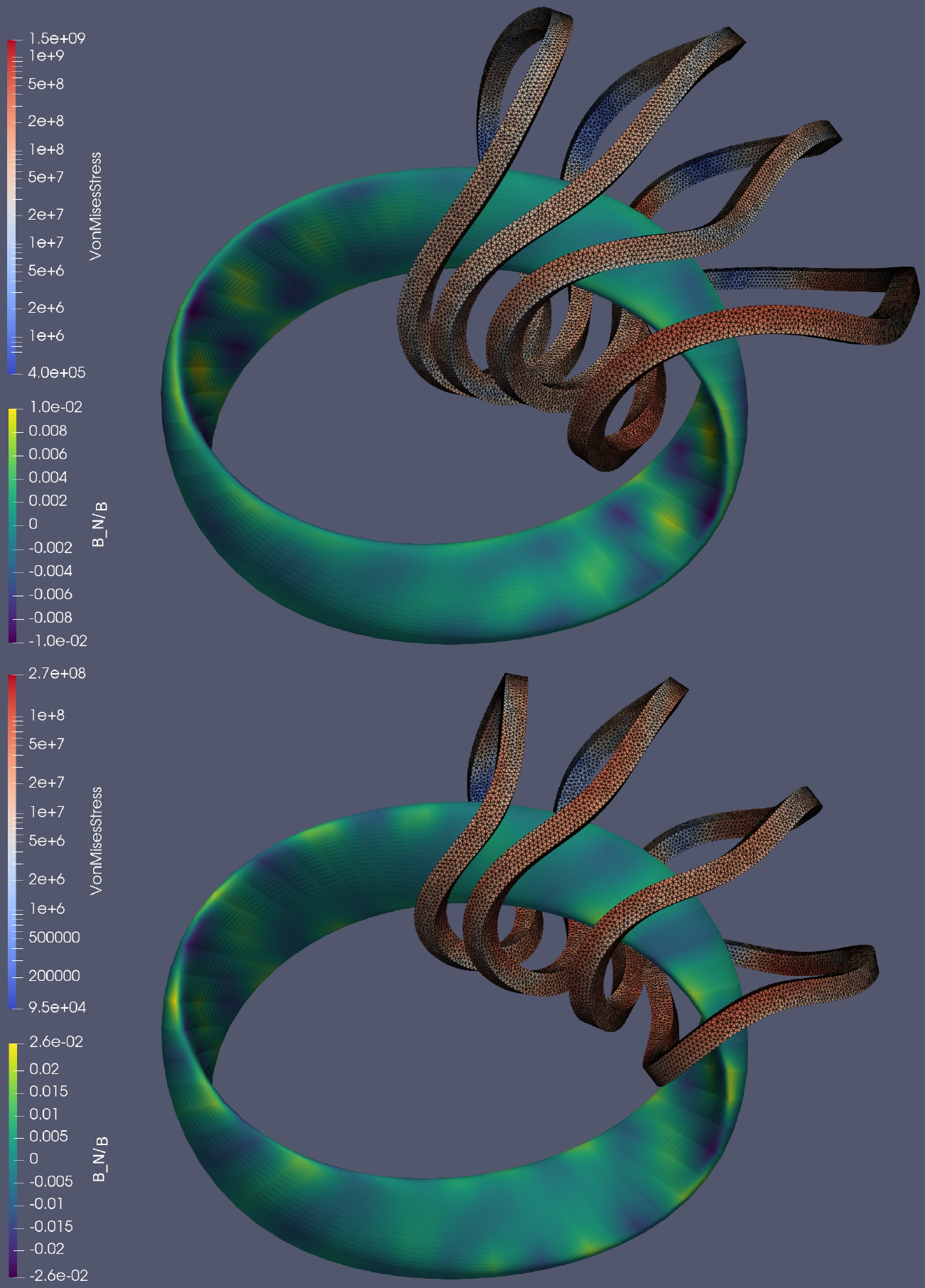}
    \caption{Left: Final coilset obtained with an order-four Landreman-Paul QA coil optimization with the default parameters. Right: Same exact optimization except there is also a penalty on volume-averaged Von Mises stress above 0.01 GPa. Note that the field accuracy decreases but the maximum Von Mises stress is reduced by about an order of magnitude.}
\label{fig:vms_optimization}
\end{figure}

Unfortunately, calculating coil displacements and Von Mises stresses requires a structural finite-element method (FEM) solve with boundary conditions specified by a predefined support structure. Such calculations for stellarator coils~\cite{packman2025bayesian,lion_stellaris_2025,swanson2025overview, biek_2026} are typically done with large commercial, i.e. closed-source, software programs like ANSYS or COMSOL. 

In order to illustrate the versatility of the code, we implemented an open-source finite-element calculation of the Von Mises stresses in the loop. As far as we are aware, this capacity has not been demonstrated before in the stellarator coil optimization. 
We minimize the volume-averaged Von Mises stress over the unique coils using a low-resolution linear elasticity solve on a tetrahedral mesh of the winding pack. Because this is a novel feature, we document the set up in detail in Appendix~\ref{sec:appendix_B}. For computational speed, the calculations are necessarily low-resolution and must be validated with the final coilset by running high-resolution, converged FEM calculations. Because we do not have a specific design point, the boundary conditions and material properties are realistic but simple in our model; more sophisticated choices will affect the absolute stress values and overall optimization problem, but should not meaningfully affect the implementation.  



A comparison of coil optimization for the Landreman-Paul QA stellarator is shown in Fig.~\ref{fig:vms_optimization} with and without a penalty on the volume-averaged Von Mises stress. The illustrated coil tetrahedral mesh is at higher resolution  for the final post-processing step than for the in-the-loop calculation. However we are not limited to low resolutions and we find that the final Von Mises stresses on a high-resolution mesh are within a factor of $\sim 2$ of the stresses reported during optimization. For the purposes of benchmarking the time taken, we also performed a fully converged run with the default Fourier continuation scheme and it ran in approximately 20 minutes using eight MPI processes on a Mac M3 MAX.

For this simple comparison in Fig.~\ref{fig:vms_optimization}, we use order-four coils and run enough iterations until the solver tolerance is achieved. The penalty is only applied if the volume-averaged stress exceeds $0.01$ GPa, which it does for the entirety of the optimization. The final coil solutions are drastically different; the optimization problem that penalizes Von Mises stresses seems to favor shorter coils but the curvature and torsion actually increase for some coils (in part because the coils are shorter). Importantly, the visualized Von Mises stresses are from the post-processing FEM calculations, showing that direct optimization during the loop has resulted in  high-resolution Von Mises stress calculations that are over an order of magnitude lower than before. Critically, the baseline coils also have unrealistic peak displacements of $\approx 16$ cm, while the peak displacement of the Von-Mises-optimized coils is only $\approx 1$ cm. This is a proof-of-concept that it is possible to directly minimize the stresses and displacements in the realistic, finite-build coil geometry.

\subsection{Genetic algorithm policy}\label{sec:genetic_algorithm}
To illustrate the capacity of the genetic algorithm and context-aware LLM policies to generate new stellarator coil designs, we present below a new coil solution for the Landreman-Paul QA stellarator that was found by the unsupervised genetic algorithm. The code submits a series of optimization batches to a supercomputing cluster. Each batch is composed according to a given policy which can be decided by an LLM or deterministically prescribed with a genetic algorithm. In the current setup, half of the batch consists of mutations of the top-$k$ successful configurations. Guardrails prevent runaway failure: if the failure rate of recent cases exceeds 60\%, for example, then the loop halts. A safe mode is activated when the failure rate exceeds 35\%, restricting the geometrically simpler plasma surfaces. 
The genetic algorithm runs with two modes:

\begin{itemize}[leftmargin=*]
\itemsep0em
    \item \textbf{Exploration}: new cases are drawn from a prior over physically meaningful parameter ranges. For exploration, we sample over the number of coils per half-field period (between 3 and 7 per half-field period), the Fourier order of the coil representation, and constraint thresholds log-uniformly distributed from reactor-scale ranges that are automatically scaled by the ratio of ARIES-CS minor radius to the chosen surface minor radius. Sampling thresholds log-uniformly creates diversity in the constraint landscape, driving the optimizer toward different local minima and effectively exploring the Pareto front between field accuracy and coil engineering feasibility. All cases are performed using the default Fourier-order continuation scheme to improve convergence. 
    \item \textbf{Exploitation}: also known as mutation, child cases are derived from a randomly chosen parent among the top-$k$ feasible configurations. For mutation, constraint thresholds are modified by multiplying a log-normal factor. Main coil parameters such as the number of coils and Fourier modes are changed to adjacent values with probability $p = 0.2$.  Weights are ignored because the augmented Lagrangian auto-tunes them; a new random seed ensures stochastic diversity even for identical configurations.
\end{itemize}

\subsection{Initial coil solutions from the genetic algorithm}\label{sec:genetic_algorithm_results}
\begin{figure}[ht]
    \centering
    \includegraphics[width=\linewidth]{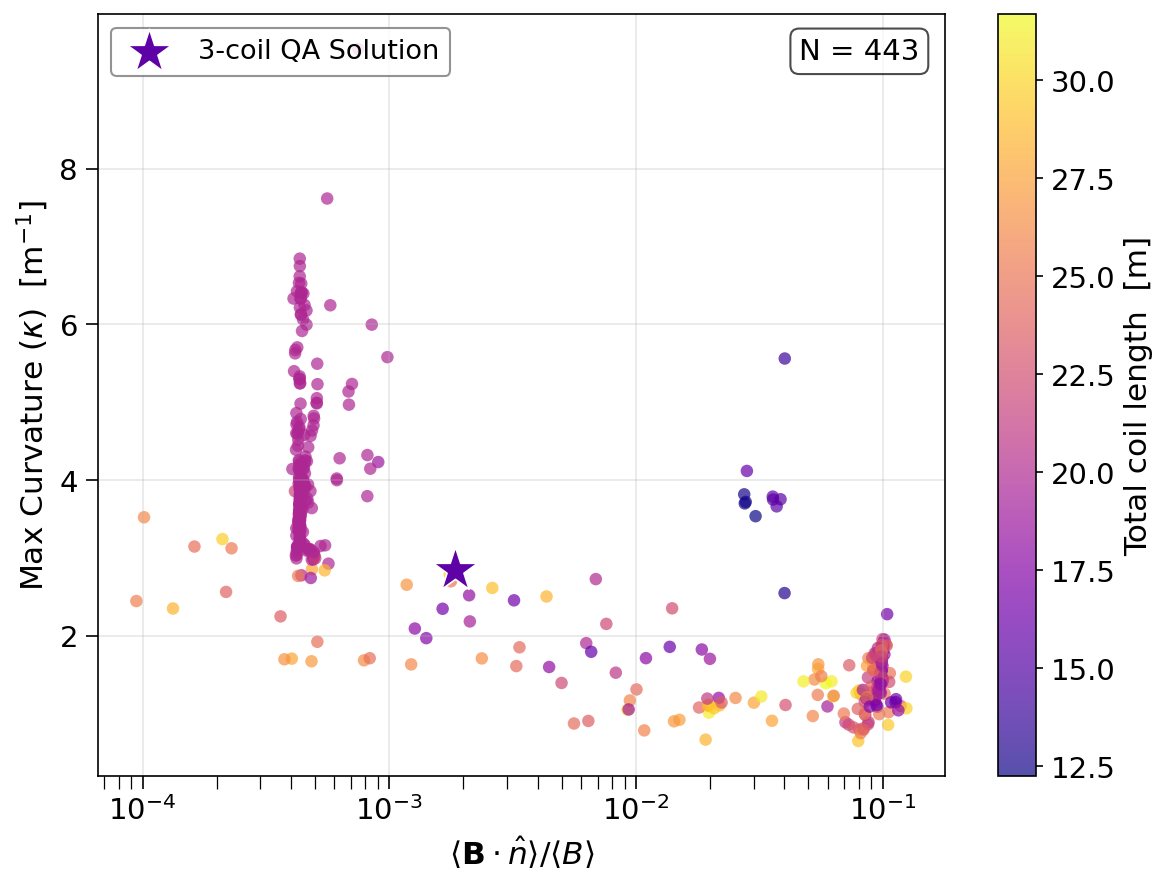}
    \includegraphics[width=\linewidth]{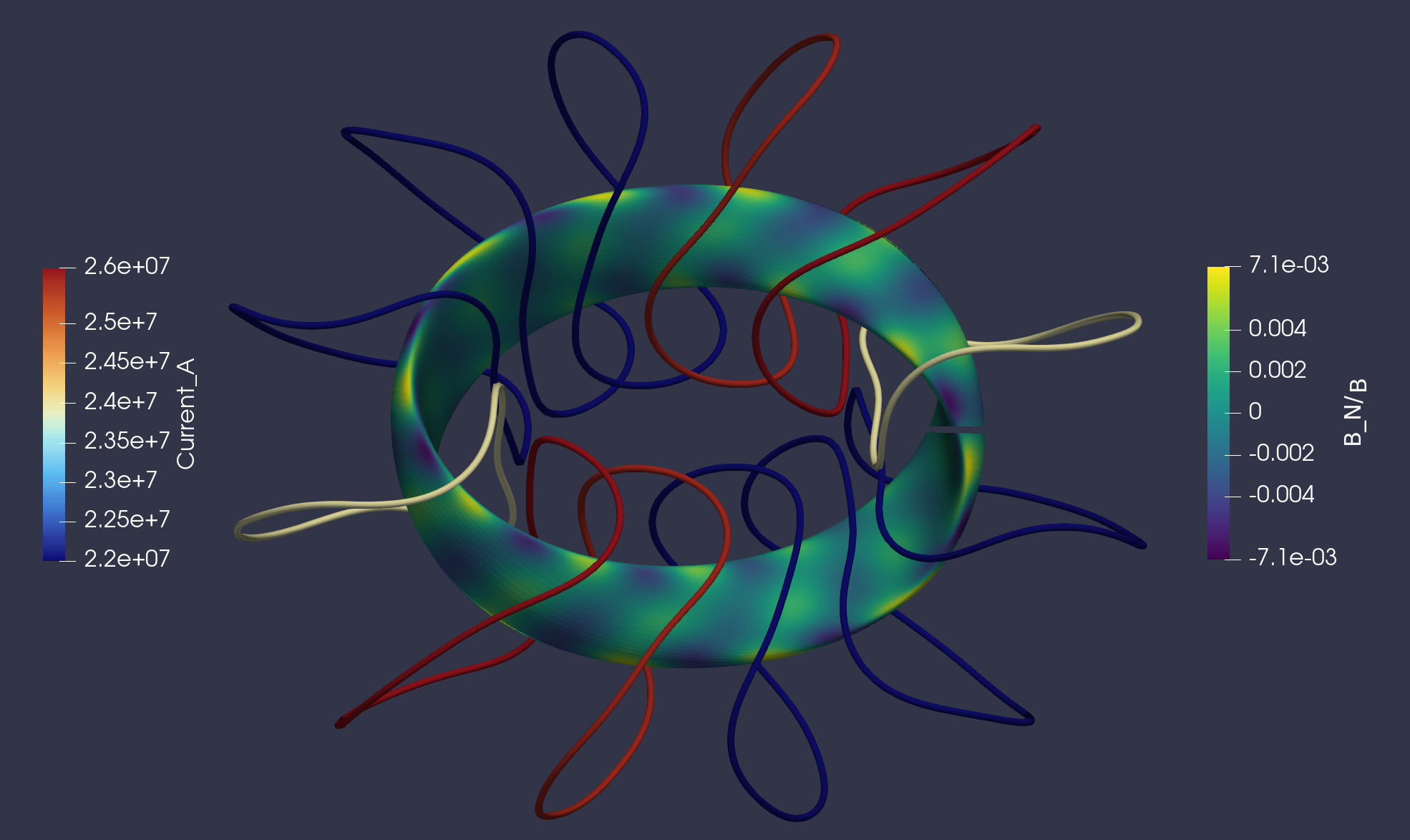}
    \caption{Top: scatter plot showing the maximum curvature versus field accuracy achieved in the 443 optimized coilsets. Note that the values plotted here correspond to the Landreman-Paul QA equilibrium scaled to 1 m major radius. The total coil length per half-field period is plotted as well. A front forms at around $\langle \bm B\cdot \hat{\bm n}\rangle / \langle B \rangle \sim 4\times 10^{-4}$ due to the imposed threshold on the squared flux. A cluster is visible at high error magnitudes, due to the exploration policy. Bottom: one of the notable configurations from this preliminary implementation of the genetic algorithm, a 3-coil per half-field period with relatively low curvature, short coils, low field errors and high coil-to-surface distance.}
    \label{fig:new_3coils_QA}
\end{figure}

Initial results yielded 443 coilsets, shown on top of Figure \ref{fig:new_3coils_QA}, where we plot the maximum curvature of the coils, a quantity for coil complexity, for each coilset against the field accuracy. We observe that most optimizations stop at $\langle \bm B\cdot \hat{\bm n}\rangle / \langle B \rangle \sim 4\times 10^{-4}$ as this corresponds approximately to the squared flux threshold set by the policy; the few configurations that manage to be more accurate are coilsets with particularly long coils. 
Note that curvatures below 2 $\text{m}^{-1}$ is already considerably restrictive for a device at 1 m major radius \cite{wechsung_precise_qs_2022, gil_augmented_2025}. Finally, a cluster at high field errors is present and is attributed to failure cases from exploration mode, this is expected behavior. Extending this database is an ongoing effort. 

Despite being limited to preliminary runs, the automated policy found a highly-competitive three-coil solution that is somewhat surprising from its low number of coils and relatively short total coil length. A three-coil solution is very advantageous from the perspective of maximizing access to the plasma for diagnostics and heating systems, but the number of coil turns to carry the enormous total current can be prohibitive. We have used the model from the Stellaris design~\cite{lion_stellaris_2025} to estimate the number of turns $(496)$ to stay below the critical current density with some margin, and the finite-build width of the coils ($0.44$ m, large but most reactor-scale coils are at least $0.3 - 0.35$m). In Table~\ref{tab:reactor_scale} we tabulate the final metrics and compare with a previous three-coil solution found by optimization in Gil et al.~\cite{gil_augmented_2025}. The total length and total superconductor needed is reduced by $\sim 17\%$; after accounting for the symmetrized coils, it requires $\sim 60$ km less superconducting material! Furthermore, it halves the max curvature, reduces the mean-squared curvature, increases the coil-surface distances, and performs comparably for minimum coil-coil distance and maximum pointwise forces. The normalized field errors worsen by roughly a factor of two but stay very good; the normalized field errors are on the scale of $\sim 0.2\%$.

\subsection{Context-aware LLM policy}

Learning optimal hyperparameters for stellarator coil optimization---thresholds, convergence tolerances, Fourier schedules, and the balance between exploit and explore---is inherently difficult.
The genetic algorithm addresses this with a fixed heuristic: log-normal mutation and log-uniform exploration.
An alternative is to use the recent explosion in the use of agentic AI models to set up an AI stellarator expert. In other words, to use a context-aware LLM policy, which reasons over the full history of runs and failures to propose targeted batches.
The LLM is used to identify under-explored plasma surfaces, avoid recurring failure modes, adjust constraint thresholds based on which margins are tight or slack, and shift between exploitation and exploration in response to score trends.
When the score trend plateaus, the policy favors exploration; when constraint violations dominate, it targets threshold adjustments to restore feasibility.
This differs from the GA not in the space of actions (both propose mutate and explore cases) but in how actions are chosen. The LLM synthesizes domain knowledge, run history, and strategic guidance rather than applying fixed sampling rules.

The policy rests on two layers of context.
First, a domain layer provides the LLM with stellarator coil optimization expertise: an optimization guide (objective terms, failure modes, metrics interpretation), threshold scaling rules at ARIES-CS reactor scale, a plasma surface catalog, and curated excerpts from a few dozen stellarator coil and equilibrium papers.
This enables informed choices even for surfaces or constraint combinations that have not yet been tried.
Second, a run-history layer summarizes the current state: top-performing runs with their scores and constraint margins, failure postmortems with rule-based suggestions, surface exploration coverage, and the score trend (improving, plateaued, or regressing).
The LLM also receives its own prior reasoning from past batches, so it can build on its choices and avoid contradicting or repeating itself.
These justifications form a running log that supports reproducibility and future analysis.
The proposer shares the same guardrails as the GA (failure-rate limits, safe mode, policy constraints); when the LLM is unavailable or returns too few valid cases, it falls back automatically to the deterministic genetic algorithm. The generation of a comprehensive LLM-generated database of solutions is ongoing and primarily limited by the need to find a consistent source of funding to support nonstop LLM calls through the APIs provided by companies such as Anthropic or OpenAI. 

\begin{table}[t]
\centering
\begin{tabular}{|c|c|c|}
\hline
Property & 3-coil QA (policy) & 3-coil QA~\cite{gil_augmented_2025} \\
\hline
Total coil length (m)                                            & $\bm{157.5}$ & 182.0  \\
Max curvature, $\kappa_{\max}$ (m$^{-1}$)                       & $\bm{ 0.28}$ & 0.53 \\
Max mean-squared $\kappa$, (m$^{-2}$) & $\bm{0.037}$ & 0.05 \\
Min coil--coil dist., $d_{cc}$ (m)                           & $\bm{1.50}$ & $1.49$ \\
Min coil--surface dist., $d_{cs}$ (m)                        & $\bm{2.95}$  &$2.74$ \\ 
$\langle \mathbf{B}{\cdot}\hat{\bm n}\rangle / \langle B\rangle$ $(\times10^{-3})$   & $1.85$ & $\bm{0.855}$  \\
Max turns for critical current & 496 & $496^*$\\
Max force (MN/m/turn) & 0.32 & $\bm{0.22}^*$\\
Linking number & 0 & 0\\
Superconductor length (km) & $\bm{75.37}$ & $90.27^*$ \\ 

\hline
\end{tabular}
\caption{Reactor-scale coil properties for the 3-coil per hfp QA coilset. Numbers are reported for the unique set of coils only. \footnotesize{$^*$For fair comparison we scaled the numbers from~\cite{gil_augmented_2025} using 496 turns and assume this number for each coil. Critical current density was not used in the previous work for determining the number of turns.}}
\label{tab:reactor_scale}
\end{table}

\section{Conclusion}\label{sec:Conclusion}
There are increasingly significant opportunities to turn over the design and optimization of fusion reactor power plants to AI tools. In particular,
the number and complexity of simulation codes for designing nuclear fusion devices is very significant and requires substantial expertise. Therefore, we have designed a new software package that can perform policy-driven automated stellarator coil optimization with built-in and extensive initialization and post-processing routines. 

In the process, we have demonstrated a new and critical ability to directly minimize Von Mises stresses in the coils from FEM calculations. This opens the door to in-the-loop optimization of many other calculations that require FEM (or calculations on a volumetric mesh), including current quench simulations, thermal analysis, and neutron fluxes.
 
Moreover, preliminary results suggest that automated policies can find competitive coil solutions without human supervision, opening the possibility to fully automating stellarator design (not just the coils, but also designing the plasma, designing other components such as the blanket, etc.) in future work. 
This would be further accelerated by differentiable codes for blanket design~\cite{bogaarts2026} and other engineering aspects. 
There remains substantial compute that must be used to build out a database of coil solutions with different plasma surfaces and perform a rigorous analysis of the performance of human optimizers versus the genetic algorithm or the context-aware LLM. 

\section{Acknowledgements}
AAK acknowledges support through the Simons Foundation under award 560651. PFG acknowledges support through the Helmholtz Association Young Investigators Group program as project VH-NG-1430. We would also like to thank Eve V. Stenson for the help supporting this project.

\appendix
\titleformat{\section}
{\normalfont\bfseries}{\appendixname\ \thesection:}{0.5em}{}
\section{Available coil objectives}\label{sec:appendix_A}
A comprehensive summary and timing benchmark of the available terms for performing coil optimization with Stellcoilbench is provided in Table~\ref{tab:coil_objective_terms}. Note that any additional terms in the primary branch of SIMSOPT are available in Stellcoilbench. The default thresholds are approximations for expected values of a reactor-scale stellarator. Initial weights on the objectives are scaled to make each objective term dimensionless. By default all quantities that depend on a length scale are rescaled by factors of $a_0 = a_\textrm{aries-cs}/a_\textrm{config}$ where $a_\textrm{config}$ is the minor radius of the chosen plasma configuration. The timing benchmark illustrates that medium-resolution FEM solves can be performed in a few seconds, and can be further sped up by using the MPI and other optimization tricks described further in the next section.

\begin{table*}[t]
\centering
\caption{Coil optimization objective terms in StellCoilBench. Benchmark: The Landreman-Paul QA stellarator with four order-eight coils; timing for objective and gradient calls is reported from an average over 10 objective and 10 gradient calls on a Mac M3 MAX. Note that the structural FEM is tested here with first-order $p=1$ basis functions but without any of the ``short-circuits'' or MPI functionality so it is a pessimistic upper bound for the scenario in which none of the optimization tricks are employed. }
\label{tab:coil_objective_terms}
\setlength{\tabcolsep}{7pt}
\adjustbox{max width=\textwidth}{%
\begin{tabular}{|c|c|c|c|c|c|c|c|}
  \hline
  Name & Term & Default Obj. & Default params. & Incl. by default? & Obj. (ms) & Grad. (ms) & Units \\
  \hline
  Squared flux & $f_b$ & $\ell_2(\lambda_{f_b})$ & $\lambda_{f_b}=10^{-8}$ & Yes & 15.6 & 51.1 & T$^2$m$^2$\\
  min coil-coil distance & $d_{\mathrm{cc}}$ & $\ell_2(\lambda_{d_{cc}})$ & $\lambda_{d_{\mathrm{cc}}}=0.8\,\mathrm{m}$ & Yes & 3.3 & 8.4 & m\\
  min coil-surface distance & $d_{\mathrm{cs}}$ & $\ell_2(\lambda_{d_{cs}})$ & $\lambda_{d_{\mathrm{cs}}}=1.3\,\mathrm{m}$ & Yes & 9.4 & 10.1 & m\\
  total coil length & $L$ & $\ell_2(\lambda_L)$ & $\lambda_L=200\,\mathrm{m}$ & Yes & 0.3 & 1.7 & m\\
  max pointwise curvature & $\kappa_{\max}$ &  $\ell_p(\lambda_\kappa)$ & $\lambda_\kappa=1\,\mathrm{m}^{-1}$, $p{=}2$ & Yes & 0.5 & 0.9 & m$^{-1}$\\
  max mean-squared curvature & $\kappa_\text{msc}$ &  $\ell_2(\lambda_{\mathrm{MSC}})$ & $\lambda_{\mathrm{MSC}}=1\,\mathrm{m}^{-2}$ & Yes & 0.6 & 1.1 & m$^{-2}$\\
  Gauss linking number & $G_L$ & --- & --- & Yes & 7.2 & 0.1 & ---  \\
  arclength variation & $V_L$ & $\ell_2(\lambda_{V_L})$ & $\lambda_{V_L}=0$ & Yes & 0.3 & 0.6 & m$^2$ \\
  max pointwise force per unit length & $F$ & $\ell_p(\lambda_F)$ & $\lambda_F=200\,\mathrm{N}/\mathrm{m}$, $p{=}2$ & No & 1.1 & 25.3 & N/m \\
  max pointwise torque per unit length & $T$ & $\ell_p(\lambda_T)$ & $\lambda_T=200\,\mathrm{N}$, $p{=}2$ & No & 1.1 & 23.6 & N \\
  pointwise torsion & $\zeta$ & $\ell_p(\lambda_\zeta)$ & $\lambda_\zeta=1\,\mathrm{m}^{-1}$, $p{=}2$ & No & 0.7 & 1.2 & m$^{-1}$ \\
  Avg. Von Mises stress ($\Delta x=0.2$m) & $\sigma_{\mathrm{vm}}$ & $\ell_p(\lambda_{\sigma_{\mathrm{vm}}})$ & $\Delta x=0.16$m, $E{=}100\,\mathrm{GPa}$ & No & 27.3 & 6660 & GPa \\
  Avg. Von Mises stress  ($\Delta x=0.1$m) & --- & --- & --- & --- & 51.2 & 9098 & --- \\
  Avg. Von Mises stress  ($\Delta x=0.05$m) & --- & --- & --- & --- & 62.6 & 13737 & --- \\
 Avg. Von Mises stress  ($\Delta x=0.01$m) & --- & --- & --- & --- & 239 & 52410 & --- \\\hline
\end{tabular}%
}
\par\vspace{0.3em}
\textbf{Algorithm default:} Augmented Lagrangian with $N_\text{AL}=10$ using L-BFGS-B with the default parameters and $N_\text{iter}=$1000.
\end{table*}

\section{Methodology for in-the-loop FEM}\label{sec:appendix_B}
FEM calculations must be optimized for speed in order to be computed in the optimization loop. We detail below a number of choices that were made to implement, make efficient, and make robust, the in-the-loop FEM calculations presented in this work.
\\ 

\textbf{Domain and Geometry:}
The coil winding pack is represented as a solid with rectangular cross-section swept along the coil centerline. The mesh is a tetrahedralization of this volume, generated either by Gmsh or a pre-built mesh from ParaStell or Gmsh. 
\\

\textbf{Mesh Deformation Strategy:}
The mesh topology is generated once and cached because the initial mesh generation is computationally expensive. On subsequent evaluations, node positions are deformed to follow the updated coil geometry:
\begin{enumerate}[leftmargin=*]
\itemsep0em
  \item For each mesh node $\mathbf{x}$, find the nearest coil centerline point $\boldsymbol{\gamma}_{i}$ and the associated right-handed centroid frame $\{\mathbf{t}_i, \mathbf{p}_i, \mathbf{q}_i\}$. Here, $\mathbf{t}_i$ is the unit tangent vector, and $\mathbf{p}_i, \mathbf{q}_i$ are unit vectors in the perpendicular plane aligned with the conductor cross-section. 
  The centroid frame is defined by~\cite{landreman2025efficient}: 
\begin{align}
    \mathbf{C} &= \operatorname{mean}(\boldsymbol{\gamma}), \quad \mathbf{w} = \boldsymbol{\gamma} - \mathbf{C},\\ \notag
    \mathbf{p} &= (\mathbf{w} - (\mathbf{w} \cdot \mathbf{t})\mathbf{t}) / |(\mathbf{w} - (\mathbf{w} \cdot \mathbf{t})\mathbf{t})|,\quad \mathbf{q} = \mathbf{t} \times \mathbf{p}.
\end{align}
  \item Next, compute the local offset in the centroid frame:
    \begin{equation}
      \delta_p = (\mathbf{x} - \boldsymbol{\gamma}_i) \cdot \mathbf{p}_i,\quad
      \delta_q = (\mathbf{x} - \boldsymbol{\gamma}_i) \cdot \mathbf{q}_i,\quad
      \delta_t = (\mathbf{x} - \boldsymbol{\gamma}_i) \cdot \mathbf{t}_i.
    \end{equation}
  \item When a coil deforms, reconstruct the new node position:
    \begin{equation}
      \mathbf{x}_{\text{new}} = \boldsymbol{\gamma}_i + \delta_p \, \mathbf{p}_i + \delta_q \, \mathbf{q}_i + \delta_t \, \mathbf{t}_i.
    \end{equation}
\end{enumerate}
This deformation preserves the local topology since each node moves with its associated centerline point, and in practice the deformations remain well-behaved enough so that the element Jacobians remain positive. Self-intersection is avoided because (1) the connectivity is fixed and (2) the offset $(\delta_p, \delta_q, \delta_t)$ is preserved, so nodes slide along the deformed centerline. For extreme curvature changes, mesh quality could degrade but in practice, we find that the various optimization tricks (see further below) and typical optimization step sizes keep the deformed mesh valid. We also found empirically that directly penalizing the curvature and torsion of the coils can help keep the mesh quality high.  
\\

\textbf{Current Density Model:}
The body force density is $\bm f = \bm J \times \bm B$, which requires a model for the current in the coil and the magnetic field and must be evaluated on the appropriate quadrature points of the weak formulation described below.
For simplicity, the current density is assumed uniform across the winding-pack cross-section:
\begin{equation}
  \mathbf{J} = \frac{I}{A} \, \mathbf{t},
\end{equation}
where $I$ is the coil current, $A$ is the cross-section area, and $\mathbf{t}$ is the unit tangent at the nearest centerline point. Each mesh node (or quadrature point) is assigned to the nearest coil centerline point via a KD-tree lookup.
\\

\textbf{Magnetic Field model:}
The Biot--Savart law diverges near the coil centerline, so a filamentary model is inadequate for points inside the conductor. StellCoilBench uses the Landreman et al.~\cite{landreman2025efficient} regularized internal field model for rectangular cross-sections. The total field at a point inside a coil is:
\begin{equation}
  \mathbf{B} = \mathbf{B}_{\text{reg}} + \mathbf{B}_0 + \mathbf{B}_{\kappa} + \mathbf{B}_b.
\end{equation}
This is a fast approximation for the self-field that significantly increases the efficiency of computing the Lorentz body force in-the-loop.
The mutual field is computed via the Biot-Savart law assuming filamentary models for the other coils, since this is adequate outside the coil where the calculations are taking place (assuming other coils do not come too close). 
\\

\textbf{Linear Elasticity:}
The linear structural problem to solve is: find the displacement $\mathbf{u} : \Omega \to \mathbb{R}^3$ such that
\begin{align}
  &-\nabla \cdot \boldsymbol{\sigma} = \mathbf{f}
  \quad \text{in } \Omega,
  \\ \notag
  \boldsymbol\sigma\mathbf{n} + K_S(\bm x)\mathbf u &= \mathbf 0
  \,\, \text{on } \Gamma_S, \,\,\, \boldsymbol\sigma\mathbf{n} = \mathbf{0}
  \,\, \text{on } \Gamma_N, \,\,\, \Gamma_S \cup \Gamma_N = \partial\Omega, \\ \notag
  K_S(\bm x) &= K_0\max\left(0, \frac{z_0 - z}{d_S}\right)^2
\end{align}
where $\mathbf n$ is the outward unit normal vector to the boundary surface $\partial\Omega$, $K_0$ is the spring modulus, $d_S$ is the depth of the support structure, $z_0$ is the end of the support structure (where $\Gamma_N$ and $\Gamma_S$ meet), and $\boldsymbol{\sigma}$ is the Cauchy stress tensor for a linear elastic, isotropic material:
\begin{equation}
  \boldsymbol{\sigma} = \lambda \, (\nabla \cdot \mathbf{u}) \, \mathbf{I} + 2\mu \, \boldsymbol{\varepsilon},
  \qquad
  \boldsymbol{\varepsilon} = \frac{1}{2}\bigl(\nabla \mathbf{u} + (\nabla \mathbf{u})^T\bigr).
\end{equation}
The Lamé Parameters are:
\begin{equation}
  \lambda = \frac{E \nu}{(1+\nu)(1-2\nu)},
  \qquad
  \mu = \frac{E}{2(1+\nu)},
\end{equation}
where $E$ is Young's modulus and $\nu$ is the Poisson ratio. The defaults are $E = 100\,\text{GPa}$, $\nu = 0.3$, which are typical for REBCO and similar materials~\cite{srivastava2024modelling}. It is straightforward to generalize the initial isotropic single-material model used here.
We have assumed a Spring-Foundation Robin boundary condition with isotropic $K(\bm x)$ with $\Gamma_S$ covering the nodes whose $z$-coordinate lies in the bottom fraction $f_S$ (default $0.15$) of each coil's $z$-range mesh:
\begin{equation}
  \Gamma_S = \bigl\{ \mathbf{x} \in \partial\Omega : z(\mathbf{x}) \leq z_{\min} + f_S(z_{\max} - z_{\min}) \bigr\}.
\end{equation}
The remainder of the boundary, $\Gamma_N$, uses the natural boundary condition. Notice that the Robin boundary condition is chosen to regularize the severity of the jump at the intersection of $\Gamma_N$ and $\Gamma_S$. 
More complex boundary conditions consistent with realistic support structures could be straightforwardly implemented in the code in future work.
\\

\begin{table*}[t]
  \centering
  \caption{The key implementation parameters for the in-the-loop optimization of Von Mises stresses. We emphasize that most of these parameters can be set to the default values.}
  \begin{tabular}{|l|l|l|}
    \hline
    Parameter & Default & Description \\
    \hline
    \texttt{width}, \texttt{height} & 0.35\,m & Cross-section dimensions \\
    \texttt{E} & 100\,GPa & Young's modulus \\
    \texttt{nu} & 0.3 & Poisson ratio \\
    \texttt{stress\_metric} & $\bar{\sigma}_{\text{vm}}$ & Scalar metric \\
    \texttt{fd\_step} & $10^{-5}$ & Finite-difference step \\
    \texttt{eval\_interval} & 1 & Evaluate every \texttt{eval\_interval} iterations \\
    \texttt{BC\_Z\_FRACTION} & 0.15 & Fraction of $z$-range for $\Gamma_S$ \\
    \texttt{structural\_quadrature\_degree} & 1 & Quadrature degree for J$\times$B\\
    \texttt{structural\_polynomial\_degree} & 2 & Order of the finite element basis\\
    \texttt{structural\_mesh\_resolution\_coarse} & 0.16\,m & Coarse mesh for adaptive strategy \\
    \texttt{structural\_mesh\_resolution\_fine} & 0.08\,m & Fine mesh after refinement \\
    \texttt{structural\_refine\_stress\_ratio} & 0.5 & Refine when $\sigma_{\text{vm}} \geq \eta \times$ threshold \\
    \texttt{structural\_use\_cached\_K} & false & Cache stiffness $\mathbf{K}$ for $\nabla J$ speedup \\
    \texttt{structural\_backend} & DOLFINx & FEM backend: DOLFINx or skfem \\
    \hline
  \end{tabular}
  \label{tab:fem_summary}
\end{table*}
\begin{figure*}[t]
    \centering
    \includegraphics[width=0.95\linewidth]{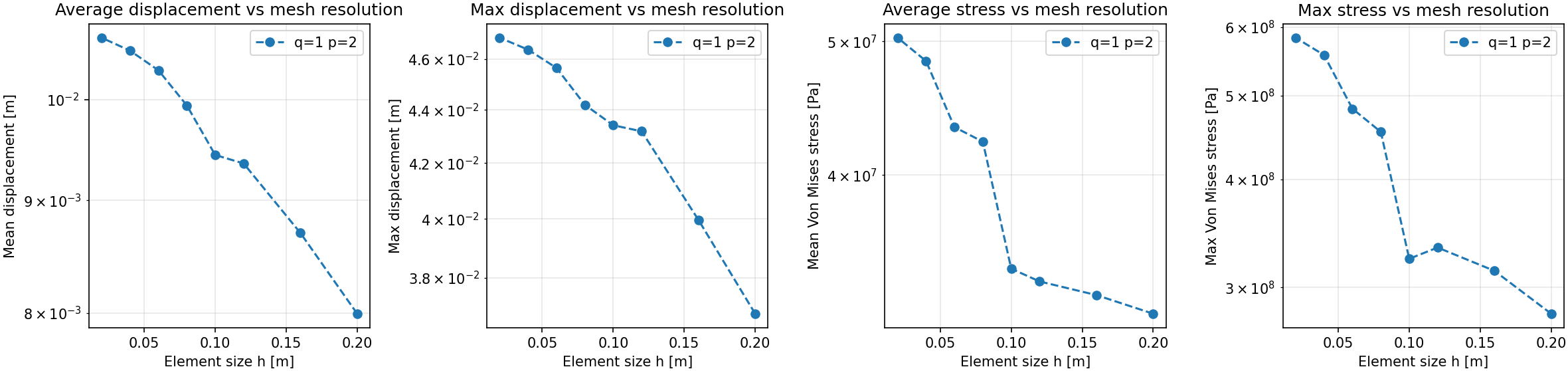}
    \caption{Convergence plots for the default settings showing (slow) convergence as the mesh resolution increases. Note the y-axes; the average displacements and average stresses do not change much in absolute terms from low to high resolution.}
    \label{fig:convergence_plots}
\end{figure*}

\textbf{Weak Form and Discretization:}
Using linear tetrahedral elements and a standard Galerkin discretization with test functions $\mathbf v$, the problem becomes:
\begin{align}
  \int_\Omega \bigl( \lambda \, (\nabla \cdot \mathbf{u}) \, &(\nabla \cdot \mathbf{v}) + 2\mu \, \boldsymbol{\varepsilon}(\mathbf{u}) : \boldsymbol{\varepsilon}(\mathbf{v}) \bigr) \, dV\\
  &= \int_\Omega \mathbf{f} \cdot \mathbf{v} \, dV - \int_{\Gamma_S} K_S(\bm x)\mathbf{u} \cdot \mathbf{v} \, dS.
\end{align}
The implementation uses scikit-fem or DOLFINx to assemble bilinear and linear forms and solve the resulting linear systems.
The weak-form right-hand side uses a quadrature with degree $q$. For $q=1$ (centroid-only), J$\times$B evaluations are reduced, giving $\sim$30--50\,\% speedup in body-force assembly with $\sim$2\,\% relative error in $J$ for typical coils found for the Landreman-Paul QA stellarator. 
%
For each tetrahedral element, the displacement gradient $\nabla \mathbf{u}$ is computed from nodal values and the element Jacobian. 
%
The deviatoric stress is $\mathbf{s} = \boldsymbol{\sigma} - \frac{1}{3}\operatorname{tr}(\boldsymbol{\sigma}) \, \mathbf{I}$. The Von Mises stress is
\begin{equation}
  \sigma_{\text{vm}} 
  = \sqrt{\frac{3}{2} \sum_{i,j} s_{ij}^2}.
\end{equation}
This is computed per element  and is constant within each linear tetrahedron.
The objective is chosen to minimize the volume-averaged stresses:
\begin{align}
  \max\left(0, \bar{\sigma}_\text{vm} - \lambda_{\sigma_{vm}}\right)^2, \quad
  \bar{\sigma}_\text{vm} = \frac{1}{V}\int \sigma_{\text{vm}}dV.
\end{align}
The Von Mises stresses use GPa units to improve optimizer numerics.
We use a number of optimization tricks to prevent undesirable behavior. 
The first is a guard against computing FEM when the coil-coil distances become very small, since the mutual field calculations will become nearly singular. When $d_{cc} < \alpha_\text{safety}\lambda_{d_{cc}}$ it skips the FEM and returns a finite penalty (e.g., 10\,GPa) with zero gradient.
The second trick is a short-circuit; when $\bar{\sigma}_\text{vm} \leq 0.9\lambda_{\sigma_{vm}}$, it skips the expensive gradient calculation and returns $\nabla J = \mathbf{0}$.
Gradients are computed by forward finite differences.
Third, the stiffness matrix $\mathbf{K}$ and the fixed degrees of freedom can be cached at the baseline; $\mathbf{K}$ changes by $\mathcal{O}(\varepsilon)$ per perturbation, which is negligible for FD accuracy.
Fourth, only coil frames and $\mathbf{B}_{\text{reg}}$ for coils whose degrees of freedom are perturbed are recomputed, reducing cost.
Fifth, the objective can be evaluated only every few iterations and cached in between to reduce cost.
Table~\ref{tab:fem_summary} summarizes the default implementation parameters for the in-the-loop FEM calculations.\\

\textbf{MPI Parallelization:}
When running with multiple MPI processes, the structural gradient $\nabla J$ is computed in parallel: rank~0 broadcasts the baseline state; each rank computes a subset of the finite-difference perturbations and partial gradients are combined.
\\

\textbf{Adaptive Mesh:}
An adaptive strategy is available in which the optimization starts with a coarse mesh and refines to a fine mesh when $\sigma_{\text{vm}} \geq \eta \lambda_{\sigma_{\text{vm}}}$  with $\eta = 0.5$ by default. Refinement is triggered from the optimizer callback; subsequent iterations use the finer mesh. This allows early iterations or low-stress regimes to run several times faster.
\\

\textbf{Dual Backends:}
DOLFINx and scikit-fem are supported but DOLFINx is typically faster per solve for in-the-loop calculations and can benefit from further parallelization. The scikit-fem package can be used if desired or as a fallback when DOLFINx is unavailable.
\\
\begin{figure}[t]
    \centering
    \includegraphics[width=0.95\linewidth]{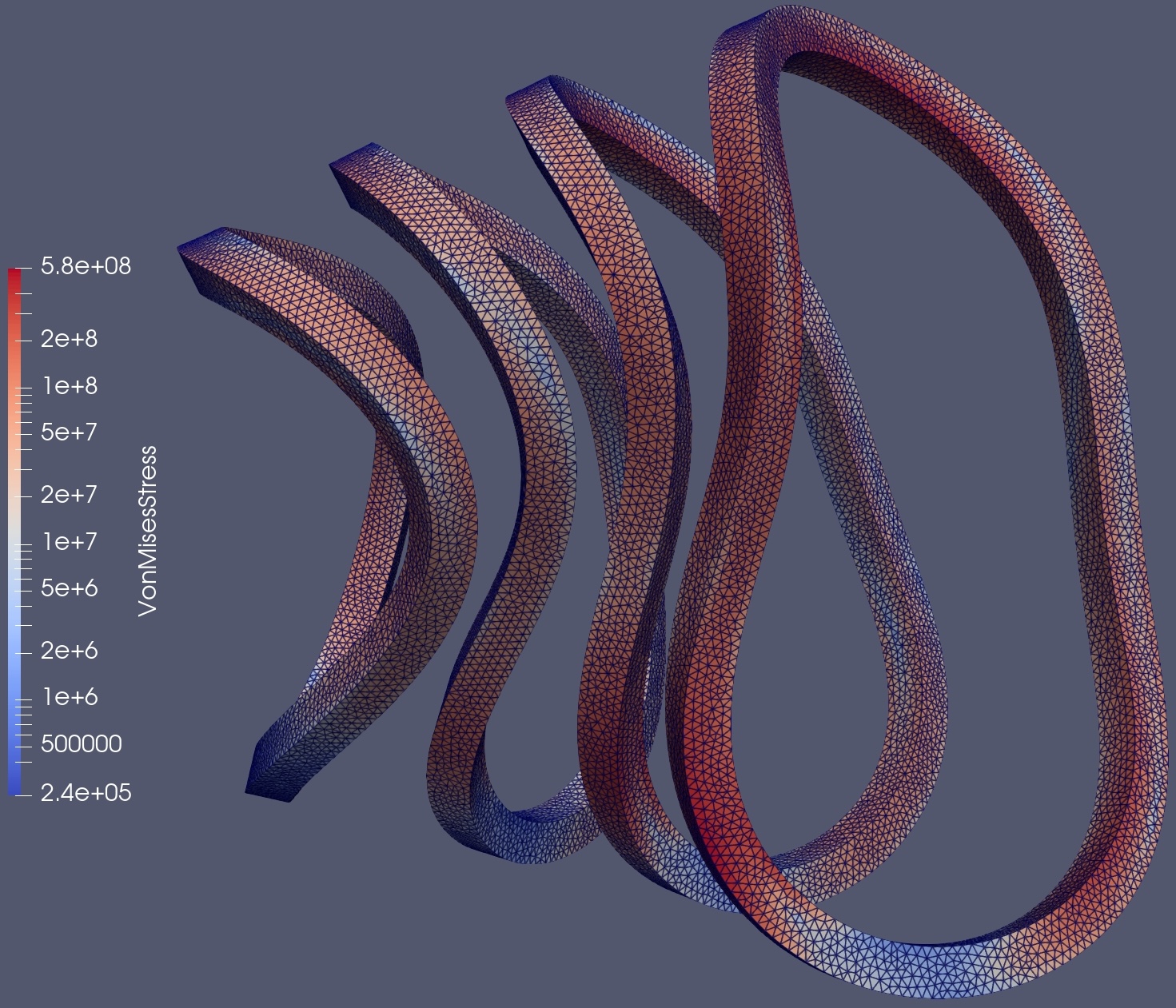}
    \caption{Log-scale of the Von Mises stresses for the highest resolution solution in the convergence test.}
    \label{fig:fem_highest_res}
\end{figure}

\textbf{Convergence tests:} Lastly, we ran some convergence tests with respect to the mesh resolution, order of the basis functions, and order of the quadrature. In Fig.~\ref{fig:convergence_plots} we show the results for average and maximum displacements and von Mises stresses for an optimized four-coil solution for the Landreman-Paul QA stellarator. The convergence trends were barely changed by going to higher order $q > 1$ quadrature or FE orders of $p > 2$ (although there is substantial benefit in $p = 2$ compared to $p=1$) so we illustrate for $(q, p) = (1, 2)$. In Fig.~\ref{fig:fem_highest_res} we illustrate the Von Mises stresses on the highest resolution grid for completeness.
The average displacements and average stresses do not change substantially in absolute terms from low to high resolution but convergence with mesh resolution is fairly slow. This behavior is well understood from the simple and artificial boundary condition used here. Singular behavior in linear elastic problems with fixed-support and Robin boundary conditions has been well-documented~\cite{sinclair2004stress}. We can also consider the limit that $K_0$ is very large where we have a fixed-support with a discontinuous jump, which develops a singularity that does not converge with mesh resolution~\cite{grisvard_elliptic_2011}. The primarily solution is to work with experimentalists to implement realistic support structure plans for reactor-scale devices. Although having in-the-loop realistic boundary conditions for the support structure is the ideal goal, this is an important first step towards a full consistent FEM integration into coil design.





\bibliography{thebibliography}

\end{document}